\documentclass[aps,prl,showpacs,floats,twocolumn,floats,superscriptaddress,floatfix]{revtex4}
\usepackage{bm}
\usepackage{times}
\usepackage{verbatim}
\usepackage{graphicx}
\usepackage{graphics,epsfig}
\usepackage{theorem}
\usepackage{makeidx}
\usepackage{amsmath}
\usepackage{epic}
\usepackage{amscd}
\usepackage{bbm}
\usepackage{xy}
\usepackage{amssymb}
\usepackage{epsfig}

\def\gsim{\;\rlap{\lower 2.5pt
\hbox{$\sim$}}\raise 1.5pt\hbox{$>$}\;}
\def\lsim{\;\rlap{\lower 2.5pt
\hbox{$\sim$}}\raise 1.5pt\hbox{$<$}\;}
\begin{document}

\newif\iffigs 
\figstrue
\iffigs \fi
\def\drawing #1 #2 #3 {
\begin{center}
\setlength{\unitlength}{1mm}
\begin{picture}(#1,#2)(0,0)
\put(0,0){\framebox(#1,#2){#3}}
\end{picture}
\end{center} }

\title{Dissipative Structures in Supersonic Turbulence}
\author{Liubin Pan}
\author{Paolo Padoan}
\author{Alexei G. Kritsuk}
\affiliation{Department of Physics, University of California, San Diego, 
CASS/UCSD 0424, 9500 Gilman Drive, La Jolla, CA 92093}  

\begin{abstract}

We show that density-weighted moments of the dissipation rate, $\epsilon_l$,
averaged over a scale $l$, in supersonic turbulence can be  
successfully explained by the She and L\'{e}v\^{e}que model [Phys.\ Rev.\ Lett.\ {\bf 72}, 336 (1994)]. 
A general method is developed to measure the two parameters of the model, $\gamma$ and $d$, based  
directly on their physical interpretations as the scaling exponent of the  
dissipation rate in the most intermittent structures ($\gamma$) and the dimension of the  
structures ($d$). 
We find that the best-fit parameters ($\gamma=0.71$ and $d=1.90$)  
derived from the $\epsilon_l$ scalings in a simulation of supersonic  
turbulence at Mach 6 agree with their direct measurements, confirming the validity  
of the model in supersonic turbulence.

\end{abstract}

\maketitle

Supersonic turbulence is ubiquitous in the cold interstellar medium~\cite{lar81} and is believed to 
play a crucial role in the process of star formation~\cite{pad02, pan081}. 
If supersonic turbulence were 
characterized by universal statistics, as often assumed for incompressible turbulence, the universality could constitute 
the foundations for a statistical theory of star formation. In this Letter we focus on the statistical properties of the most 
intermittent structures (MISs) of supersonic turbulence.

The theory of fully developed turbulence assumes that the scaling behavior of small-scale fluctuations in the 
inertial range is flow-independent, e.g., the moments of the velocity difference, 
$\langle \delta v(l)^p \rangle =  \langle (v(x+l)-v(x))^p \rangle \propto l^{\zeta_p}$, 
have universal scaling exponents, $\zeta_p$. The universal state for fully developed incompressible 
turbulence proposed by Kolmogorov in 1941 (K41)~\cite{kol41}, with $\zeta_p = p/3$, has been shown to deviate 
significantly from 
$\zeta_p$ measured in both experiments and numerical simulations, at $p>3$. 
This discrepancy is due to spatial fluctuations in the dissipation rate, neglected in the K41 theory 
\cite{lan44}. 
The scaling exponents of the average energy dissipation, $\epsilon_l$, over a scale $l$ (see eq. (3)) 
give corrections to the K41 theory, referred to as intermittency corrections~\cite{kol62, she94}. 
A careful study of fluctuations in the energy dissipation is essential for understanding intermittency in turbulence.

The intermittency model by She and L\'{e}v\^{e}que~\cite{she94} (hereafter the SL model) is based on a hierarchy of 
dissipative structures of different intensity levels, 
characterized by the ratios, $\epsilon_l^{(p)}= \langle \epsilon_l^{p+1} \rangle/\langle \epsilon_l^{p} \rangle$, of successive moments 
of $\epsilon_l$. With increasing order, $p$, $\epsilon_l^{(p)}$ represents structures of increasing intensity and $\epsilon_l^{(\infty)}$ 
corresponds to the MISs.
By invoking a hypothetical ``hidden symmetry" that relates this hierarchy of structures 
to the most intermittent ones, the model predicts the scaling exponents, $\tau_p$, of the energy dissipation moments, 
$\langle \epsilon_l^p \rangle \propto l^{\tau_p}$, of all orders, $p$,
\begin{equation}
\tau_p=-\gamma p + \gamma (1-\beta^p)/(1-\beta). 
\label{eq1}
\end{equation}
The parameter $\gamma$ is the scaling exponent of the dissipation rate in the MISs, 
$\epsilon_l^{(\infty)} \propto \lim_{p \to \infty} l^{\tau_{p+1}-\tau_p} \propto l^{-\gamma}$, and $\beta$ is related to $\gamma$ 
and to the Hausdorff dimension, $d$, of the MISs by $\gamma/(1-\beta) = D-d$, where $D=3$ for three-dimensional (3D) turbulence. 
The physical meaning of this relation will be explained later. This model is very successful in predicting  $\zeta_p$ in incompressible turbulence with high accuracy.   

In this Letter, we study the fluctuations of the dissipation rate in supersonic hydrodynamic (HD) 
turbulence using numerical simulations. We show that the simulation results for the scaling exponents, 
$\tau_p$, are well represented by eq. (1), suggesting the SL 
formulation for the scaling behavior of the dissipation rate, originally proposed 
for incompressible turbulence, may be applied to supersonic turbulence as well. 
We present a method to directly measure the parameters $\gamma$
and $d$ according to their physical interpretation, which is general  
and not limited to the supersonic regime of interest here. This method can be  
used to test the validity of the physical interpretation of the SL model in any of  
its applications. For supersonic turbulence, we find that the 
parameters  derived with this method are in excellent agreement with 
the values that best fit $ \tau_p$, which confirms the physical interpretation of the model.



Instead of directly investigating the statistics of the dissipation rate, most studies of this model 
are primarily concerned with the scaling exponent, $\zeta_p$, of the velocity difference, $\delta v(l)$. The model predicts 
$\zeta_p= (1-\gamma)p/3 + \gamma (1-\beta^{p/3})/(1-\beta)$,
which follows from the refined similarity hypothesis, $\delta v(l) \sim \epsilon_l^{1/3} l^{1/3}$, and eq (1). Assuming that the largest 
available kinetic energy in the strongest structures is $\sim U^2$, with $U$ being the rms velocity, and that the timescale in these structures 
follows the usual Kolmogorov scaling, $t_l \propto l^{2/3}$,  She and L\'{e}v\^{e}que argued that $\epsilon^{(\infty)} \sim U^2/t_l \propto l^{-2/3}$, 
i.e., $\gamma=2/3$\footnote{A direct check for the accuracy of this argument 
can be done with our method for measuring $\gamma$.}. With this $\gamma$ and with $d=1$, corresponding to filamentary dissipative structures, the values of
$\zeta_p$ predicted by this model agree with experimental results of incompressible turbulence with an accuracy of about 1\%~\cite{she95}.

Although not directly measured from the MISs, 
(cf.~\cite{she01}), 
$\gamma=2/3$ has been adopted in most applications of the model to 
incompressible~\cite{mul00} and supersonic MHD turbulence~\cite{bol02, pad04}. The dimension $d$ was obtained either 
from the assumed geometry of the MISs ($d=2$ for current sheets or shocks in MHD or supersonic turbulence), or from the best fit to the 
numerical velocity structure functions~\cite{pad04, kri071}. These works have shown that the SL model with 2D 
MISs is generally consistent with 
simulations of MHD and highly compressible turbulence.

However, a strict verification of the validity of this model requires a demonstration that the parameters that fit $\tau_p$ 
also have the declared physical meaning, otherwise the agreement between the model and the simulations may be a 
mere coincidence. In the present work we thus obtain $\gamma$ and $d$ both by a direct measurement from their interpretations, and by 
fitting $\tau_p$. We are also interested in deriving the dimension of the MISs in supersonic 
HD turbulence where about 1/3 kinetic energy dissipates in dilatational modes and 2/3 in solenoidal modes, 
with strongest shocks generally coinciding with the locations of strongest shear and vortices. 


{\it Measuring $\tau_p$.}--We take the 1024$^3$ simulation of supersonic HD turbulence for isothermal ideal gas 
with a rms Mach number of 6 from reference~\cite{kri072}. The simulation employs the piecewise parabolic method to solve 
the Euler equation~\cite{col84}. We focus on the statistics of $\epsilon_l$. The dissipation rate per unit mass at a 
given position and time is calculated by~\cite{lan87},  
\begin{equation}
\epsilon ({\bf x},t)
= (2 {\rm Re})^{-1} (\partial_i v_j + \partial_j v_i -(2/3)\delta_{ij} \partial_k v_k) ^2      
\label{eq1}
\end{equation}
where $Re$ is the effective Reynolds number controlled by numerical dissipation. 
We compute the velocity gradients at the resolution scale and assume $Re$ is constant. 
 
We calculate the average dissipation rate, $\epsilon_l({\bf x}, t)$, at a scale $l$ around ${\bf x}$,  
from the definition given in~\cite{kol62} (generalized to account for density fluctuations), 
\begin{equation}          
\epsilon_l ({\bf x},t) = \frac{1}{{\rho}_l({\bf x},t)V(l) } \int\limits_{|{\bf x'}|<l} \rho( {\bf x}+{\bf x'},t ) 
\epsilon ({\bf x}+{\bf x'},t) d{\bf x'}  
\label{eq3}
\end{equation} 
where $V(l)=4 \pi l^3/3$ is the volume of a spherical region of size $l$, and  
${\rho}_l ({\bf x},t) = 1/V(l) \int_{ |{\bf x'}|<l } \rho( {\bf x}+{\bf x'},t ) d{\bf x'}$ is the average density 
of that region. For convenience, we divide the simulation box into cubes (instead of spheres) of different sizes in our computations.

The moments, $\langle \epsilon_l^p \rangle$, of  $\epsilon_l$, can be evaluated by  
\begin{equation}          
\langle \epsilon_l^p \rangle  = \frac{1} { {\bar \rho} V} \int \epsilon_l^p({\bf x},t)\rho_l( {\bf x},t ) d{\bf x} 
\label{eq4}
\end{equation} 
where $V$ is the total volume of the system and ${\bar \rho}$ is the overall average density. We have used the Favre~\cite{fav58} density 
weighting factor, $\rho_l/{\bar \rho}$, 
to account for the density variations in compressible turbulence. 
With this density weighting, the first order moment, i.e., $\langle \epsilon_l \rangle$, is independent of 
scale $l$, as follows from eqs. (3) and (4), resulting in $\tau_1=0$ (see Fig. 1). This suggests that, if the refined similarity hypothesis applies 
to supersonic turbulence, the density-weighted third-order velocity structure function in compressible 
turbulence would have $\zeta_3=1$, an exact result for incompressible turbulence, known as Kolmogorov's 4/5 law~\cite{kri072}.
Since the SL formula, eq. (1), gives $\tau_1=0$, it is appropriate to compute the density-weighted moments from the  
simulation data and compare with the model. The Hausdorff dimension of the MISs 
we obtain from fitting $\tau_p$ is thus density-weighted in the sense of density-weighting in eq. (4). 
For a valid comparison, we will include density-weighting in our direct measurement of $d$.       
     
We calculate moments of $\epsilon_l$ from eq. (4) for 9 snapshots of our simulation, covering more than 5 dynamical times. We obtain 
$\tau_p$ from least-square fits to the $\log_{10}(\langle \epsilon_l^p \rangle)$-$\log_{10}(l)$ curves in each snapshot.
The results are shown in Fig. 1, where the data points and the error bars are, respectively, the average exponents and the standard 
deviations over the 9 snapshots. The error bars are negligible for $p<2$, meaning that there are little snapshot-to-snapshot variations 
for the exponents at low orders. The scatter increases with the order, and the error bar at $p=4$ is already significant (7\%).  
We find that, starting from $p=4$, the $\log_{10}(\langle \epsilon_l^p \rangle)$-$\log_{10}(l)$ curves are no longer well fit by 
straight lines and thus we only show results up to the 4th order. Note, however, that the 4th order moment of $\epsilon_l$ corresponds to 
12th order moment of $\delta v(l)$.

\begin{figure}
\includegraphics[width=0.455\textwidth]{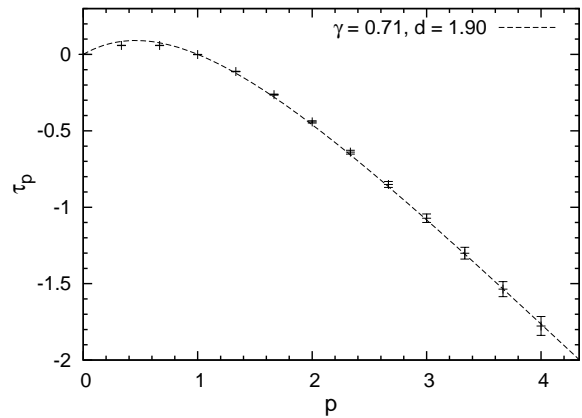} 
  \caption{Scaling exponents, $\tau_p$, of the dissipation rate, $\epsilon_l$, averaged over
9 snapshots. Error bars indicate snapshot-to-snapshot variations. The best fit gives $\gamma=0.71$ and $d=1.90$. }
\label{tau}
\end{figure}

Comparing with eq. (1), we find that the SL model with $\gamma=0.71$ and $\beta=0.35$ ($d=1.90$) gives an excellent fit to the numerical data. 
The fit shows that the SL model can be successfully 
applied to the density-weighted statistics of the dissipation rate in supersonic turbulence. 
As mentioned earlier, a demonstration that the best-fit parameters indeed carry their physical meaning is needed 
to verify the validity of the model. To this end, we directly measure the parameters from the simulation data. 
A fairly large range of parameter pairs, (0.67-078) for $\gamma$, and a corresponding range of (2.04-1.60) for $d$, 
can give acceptable (but poorer) fits to the numerical results for $\tau_p$ 
within the 2 $\sigma$ error bars. If the SL model works, a direct measurement would fix these parameters.

{\it Measuring $\gamma$.}--
We obtain $\gamma$ directly by measuring the average dissipation 
rate profile around the MISs.  
We first locate the MISs
in the computational domain, by selecting cells with 
dissipation rate larger than a given threshold, $\epsilon_{th}$, set to be 
close to 
the maximum 
dissipation rate over the domain, $\epsilon_{m}$. We will call these cells the dissipation peaks. 
We then use cubic boxes of different sizes, $l$, to cover each peak, and evaluate the average dissipation rate in each 
box, $\epsilon_p(l)$, through eq. (3) (with ${\bf x}$ at the peak). Taking the average over all 
peaks, we obtain an average dissipation rate, $\langle \epsilon_p(l) \rangle$, as a function of 
the cube size, $l$. The slope of the profile is expected to be the exponent, $\gamma$. 
We calculated the average $\epsilon_p(l)$ in two different ways: with and without the average 
density $\rho_p(l)$ in a box of size $l$ as a weighting factor. We find little difference between the 
slopes obtained in the two ways, implying a weak correlation between the dissipation rate and the density around the 
peaks \footnote{
We do not use density weighting for $\langle \epsilon_p(l) \rangle$ shown in Fig. 2. It 
is more appropriate to use it in the probability of finding MISs 
at each scale when measuring $d$. 
}. 
We increase the threshold and check whether the slope of $\epsilon_p(l)$ converges. 
The converged slope is the parameter $\gamma$ that we pursue.

\begin{figure}
\includegraphics[width=0.45\textwidth]{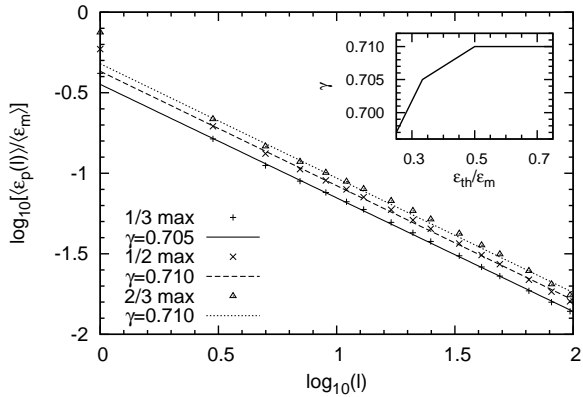}
 \caption{Average dissipation rate, $\langle \epsilon_p(l) \rangle$, (normalized to 
average maximum, $\langle \epsilon_{m} \rangle$) around dissipation peaks 
selected by thresholds, 1/3, 1/2 and 2/3 $\epsilon_{m}$. 
Best-fit lines give $\gamma=0.705$, $0.710$ and $0.710$, respectively.  
The inset shows convergence of measured $\gamma$ with $\epsilon_{th}$.
\label{Gamma}}
\end{figure}

Our result is shown in Fig. 2 for three thresholds, 1/3, 1/2 and 2/3 $\epsilon_{m}$. The three curves are the 
average over all the peaks in the same 9 snapshots used to calculate $\tau_p$. 
The profiles are approximated well by power laws (except at $l=1$, i.e., at the resolution scale) and we find 
$\gamma=0.705$, $0.710$ and $0.710$, respectively, for the three thresholds. This value is very close to 2/3 
proposed by She and L\'{e}v\^{e}que, suggesting that the Kolmogorov scaling for the timescale in the MISs, 
$t_l \propto l^{2/3}$, applies also to supersonic turbulence. 
The measured $\gamma$ converges to 0.71 at the threshold of 1/2 $\epsilon_m$, which concides with 
the value obtained from the best fit to $\tau_p$.   
Besides showing the applicability of the SL model to supersonic turbulence, we have thus verified that the 
parameter $\gamma$ carries the precise physical meaning in the model.

{\it Measuring $d$.}--The Hausdorff dimension, $d$, enters the SL model through the following argument, 
which also provides an explanation of the relation $\gamma/(1-\beta)=D-d$. In the limit $p \to \infty$, the contribution to 
$\langle \epsilon_l^p \rangle$ would be primarily from the MISs at scale $l$. Since the average 
dissipation rate in regions of size $l$ containing MISs goes like $l^{-\gamma}$, we have 
$\langle \epsilon^p_l \rangle \propto l^{-\gamma p} P(l)$, in the limit $p \to \infty$, where $P(l)$ is 
the probability of finding a region of size $l$ that hosts a dissipative structure of the highest level at scale $l$. 
A geometric consideration suggests that $P(l) \propto l^{D-d}$ if the dimension of the MISs is $d$~\cite{fri95}. 
This gives $\langle \epsilon_l^{p} \rangle \propto l^{-\gamma p + D-d}$ as $p \to \infty$. It immediately follows from eq. (1) that 
$\gamma/(1-\beta)=D-d$. 

\begin{figure}
\includegraphics[width=0.45\textwidth]{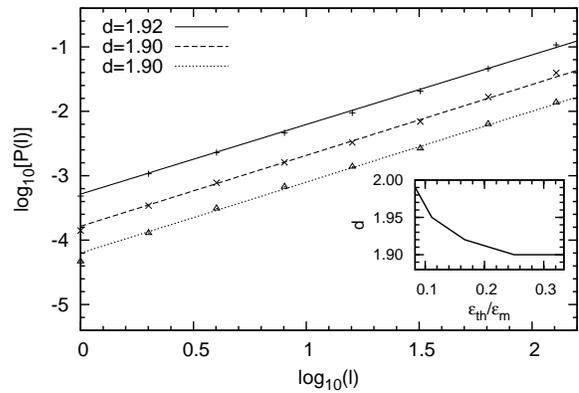}
  \caption{Probability of finding cubes of size $l$ with average dissipation rate larger than $1/6$, $1/4$ and $1/3$ $\epsilon_{m} l^{-0.71}$. 
Scaling exponents of $P(l)$ correspond to $d=1.92$, $1.90$ and $1.90$, respectively. The inset shows convergence of measured $d$ with 
$\epsilon_{th}$.}         
\label{Dimension}
\end{figure}

Directly measuring the Hausdorff dimension of the MISs, e.g., using a box-counting method, is challenging. 
Here we take a simpler approach: we compute the probability, $P(l)$, of finding an MIS in a cube of size $l$,  
and derive $d$ from the scaling of $P(l)$ with $l$, based on the physical 
argument given above
We need a criterion to judge whether a cube of size $l$ in the simulation box contains an MIS at that scale. Based on the 
log-Poisson version of the SL model~\cite{dub94, she95}, we find that the appropriate criterion is that the cube in question 
has an average dissipation rate larger than a threshold that scales like $\epsilon_{th} l^{-\gamma}$ with $l$. The factor 
$l^{-\gamma}$ accounts for the decrease of the average dissipation rate in the MISs with scale. 
We will let $\epsilon_{th}$ 
approach $\epsilon_{m}$ in the simulation box. 

The chosen threshold is justified as follows. The SL model is equivalent to a log-Poisson distribution for 
$\epsilon_l$, i.e.,  $P(\epsilon_l)d\epsilon_l = (l/L)^{(D-d)} \Sigma_{n=0}^{\infty} \frac {\lambda^n}{n!} P_L(\ln(\epsilon_l l^{\gamma} ) - \ln({\bar \epsilon} L^{\gamma})-n \ln(\beta) )d \ln(\epsilon_l)$ where $L$ is the integral scale, ${\bar \epsilon}$ the overall dissipation rate, 
and $\lambda=(D-d)\ln(L/l)$~\cite{pan081, pan082}. The distribution, $P_L(\ln(\epsilon_L/{\bar \epsilon}))$, of the dissipation rate, 
$\epsilon_L$, at $L$, depends on the driving force and is thus non-universal. At the integral scale, $\epsilon_L$ is approximately 
equal to ${\bar \epsilon}$, thus $P_L(x)$ is supposed to be narrow and decrease very rapidly with increasing $x$. 
If the SL model is valid for supersonic turbulence, the probability, $P(l)$, 
of finding a region of size $l$ with $\epsilon_l> \epsilon_{th} l^{-\gamma}$ is given by the cumulative probability $P(l)=(l/L)^{(D-d)} 
\Sigma_{n=0}^{\infty} \frac {\lambda^n}{n!} \int_{\ln(\epsilon_{th}/{\bar \epsilon})-\gamma \ln (L) -n \ln(\beta)} P_L(x) d x$ 
where $l$ and $L$ are in units of the resolution scale. Due to the rapid decrease of $P_L(x)$ with $x$, for a large $\epsilon_{th}$ 
the contribution from the $n$-th term to $P(l)$ decreases quickly with $n$ because the lower integral limit increases 
with $n$ (since $\beta<1$). As $\epsilon_{th}$ increases, the contribution would be more and more 
dominated by the $n=0$ term, which is $\propto l^{(D-d)}$. 
Therefore, in our measurement, we increase $\epsilon_{th}$ and check whether the scaling 
exponent of $P(l)$ converges. If the SL model is correct, the converged exponent is expected to be the 
codimension $D-d$ of the MISs and to agree with that derived from $\tau_p$. To be consistent with the 
density-weighting in the dissipation rate moments, each cube that satisfies the criterion is given a weighting factor proportional to 
the average density in the cube.

Fig. 3 shows the scaling of $P(l)$ with $l$ (averaged from the 9 snapshots), 
for 3 different $\epsilon_{th}$: $1/6$, $1/4$, and $1/3$ $\epsilon_{m}$. 
For the 3 choices of $\epsilon_{th}$, the scaling exponents of $P(l)$ are, respectively, 1.08, 1.10, and 1.10 (meaning $d=1.92$, $1.90$ and $1.90$). 
As $\epsilon_{th}$ increases, $d$ converges to 1.90, which again agrees exactly with the best-fit value 
from the scaling of the dissipation rate moments.
Together with the measurement of $\gamma$, this estimate of $d$ 
validates the extension of the SL model to supersonic turbulence, and confirms the validity of the 
physical interpretation of the parameters.
               
In conclusion, we have studied the statistics of energy dissipation in supersonic HD turbulence at Mach number $M=6$ using numerical simulations. 
We have computed the scaling exponents, $\tau_p$, of density-weighted moments of the dissipation rate, $\epsilon_l$, 
averaged over a scale $l$, and found that the SL intermittency model with $\gamma=0.71$ and $d=1.9$ gives an excellent fit to the measured $\tau_p$. 
We stress that, with density weighting, $\tau_1=0$, suggesting a linear scaling for the density-weighted 3rd order velocity structure function. 
We have developed a general method to directly measure $\gamma$ and $d$, which provides a validity test of the 
physical interpretation of the model. We have shown that the parameters measured directly are exactly equal to the values that best fit $\tau_p$. 
We have thus verified that the SL model can be successfully applied to supersonic turbulence. 
Investigations with other Mach numbers, especially larger ones, would advance our 
understanding of the energy dissipation in supersonic turbulence. At large enough $M$, there may exist an asymptotic state 
(possibly already reached at $M=6$), where the scaling of the energy dissipation rate (i.e., $\tau_p$) 
would be universal and independent of $M$, and so would $\gamma$ and $d$. 
This conjecture is based on the observation that, at $M \ge 6$, an equilibrium in kinetic energy partition 
between the solenoidal modes (2/3) and the potential modes (1/3) is always established (for an isothermal equation of state), 
regardless of their energy ratio in the driving force.

This research was partially supported by a NASA ATP grant NNG056601G, by NSF grants AST-0507768, AST-0607675 
and NRAC allocation MCA07S014. We utilized computing resources provided by the San Diego Supercomputer
Center, by the National Center for Supercomputing Applications and by NASA High End Computing Program.

Note added in proof.--The authors are grateful to Dr. Chris McKee for pointing out an error in Eq.(2) in an early draft 
of this Letter.


\begin{thebibliography}{60}
\bibitem{lar81}
R. B. Larson, Mon. Not. Roy. Astron. Soc., {\bf 194}, 809, 1981;
M. H. Heyer and C. M. Brunt, ApJ, {\bf 615}, L45, 2004. 
P. Hily-Blant, E. Falgarone and J. Pety, A\&A, {\bf 481}, 367, 2008
\bibitem{pad02}
P. Padoan and A. Norlund, ApJ, {\bf 576}, 870, 2002; 
P Padoan et al., ApJ, {\bf 661}, 972, 2007
\bibitem{pan081}
L. Pan and P. Padoan, astro-ph/0806.4970, 2008.
\bibitem{kol41}
A. N. Kolmogorov, Dokl, Akad, Nauk SSSR, {\bf 30}, 301, 1941
\bibitem{kol62}
A. N. Kolmogorov, J. Fluid Mech. {\bf 13}, 82, 1962
\bibitem{lan44}
The remark by Laudau presented in a scientific discussion was incorporated in the 
first edition (in Russian) of the book: L. D. Landau and E. M. Lifshitz, 
Mechanics of Continuous Media (Gostechnicisdat, Moscow,  1944)
\bibitem{she94}
Z-S. She and E. Leveque, Phys. Rev. Lett, {\bf 72}, 336, 1994
\bibitem{she95}
Z-S. She and E. C. Waymire, Phys. Rev. Lett, {\bf 74}, 262, 1995
\bibitem{she01}
Z-S. She, K. Ren, G. S. Lewis, and H. L. Swinney, Phys. Rev. E, {\bf 64}, 016308, 2001
\bibitem{mul00}
W-C. Muller and D. Biskamp, Phys. Rev. Lett., {\bf 84}, 475, 2000; 
J. Cho, A. Lazarian, and E. T. Vishniac, ApJ, {\bf 564}, 291, 2002
\bibitem{bol02}
S. Boldyrev, ApJ, {\bf 569}, 841, 2002
\bibitem{pad04}
P. Padoan, R. Jimenez, A. Nordlund, and S. Boldyrev, Phys. Rev. Lett., {\bf 92}, 191102, 2004
\bibitem{kri071}
A. G. Kritsuk, P. Padoan, R. Wagner, and M. L. Norman, AIPC, {\bf 932}, 393, 2007
\bibitem{kri072}
A. G. Kritsuk, M. L. Norman, P. Padoan, and R. Wagner, ApJ, {\bf 665}, 416, 2007
\bibitem{lan87}
L. D. Landau and E. M. Lifshitz, Fluid Mechanics (Pergamon Press, 1987), pg. 194, eq. (49.5). 
\bibitem{fav58}
C. R. Favre, {\it Acd. Sci., Paris, Ser. A}, {\bf 246}, 2576, 1958
\bibitem{col84}
P. Colella and P. R. Woodward, J. Comp. Phys., {\bf 54}, 174, 1984; I. V. Sytine et al., J. Comp. Phys., {\bf 158}, 225, 2000 
\bibitem{fri95} 
U. Frisch, Turbulence, Cambridge University Press, 1995
\bibitem{dub94} 
B. Dubrulle, Phys. Rev. Lett., {\bf 73}, 959, 1994
\bibitem{pan082} 
L. Pan, J. C. Wheeler, and J. Scalo, ApJ, {\bf 681}, 470, 2008 
\end{thebibliography}
\end{document}